# Structural characterization of $SiO_2$-$Na_2O$-$CaO$-$B_2O_3$-$MoO_3$ glasses

D. Caurant, O. Majérus, E. Fadel, M. Lenoir
*CNRS, ENSCP, Laboratoire de Chimie de la Matière Condensée de Paris*
*(UMR-CNRS 7574), 75231 Paris, France*

C. Gervais
*CNRS, Université Pierre et Marie Curie, Laboratoire de Chimie de la Matière Condensée de Paris (UMR-CNRS 7574), 75252 Paris, France*

T. Charpentier
*CEA Saclay, Laboratoire de Structure et Dynamique par Résonance Magnétique,*
*DSM/DRECAM/SCM-CEA/CNRS URA 331, 91191 Gif sur Yvette, France*

D. Neuville
*Laboratoire de Physique des Minéraux et Magmas CNRS-IPGP, 4 Place Jussieu,*
*75252, France*

Nuclear spent fuel reprocessing generates high level radioactive waste with high Mo concentration that are currently immobilized in borosilicate glass matrices containing both alkali and alkaline-earth elements [1]. Because of its high field strength, $Mo^{6+}$ ion has a limited solubility in silicate and borosilicate glasses and crystallization of alkali or alkaline-earth molybdates can be observed during melt cooling or heat treatment of glasses [2-4]. Glass composition changes can significantly modify the nature and the relative proportions of molybdate crystals that may form during natural cooling of the melt. For instance, in a previous work we showed that $CaMoO_4$ crystallization tendency increased at the expenses of $Na_2MoO_4$ when $B_2O_3$ concentration increased in a $SiO_2$-$Na_2O$-$CaO$-$MoO_3$ glass composition [4]. In this study, we present structural results on two series ($M_x$, $B_y$) of quenched glass samples belonging to this system using $^{29}Si$, $^{11}B$, $^{23}Na$ MAS NMR and Raman spectroscopies. The effect of $MoO_3$ on the glassy network structure is studied and its structural role is discussed ($M_x$ series). The evolution of the distribution of $Na^+$ ions within the borosilicate network is followed when $B_2O_3$ concentration increased ($B_y$ series) and is discussed according to the evolution of the crystallization tendency of the melt. For all glasses, ESR was used to investigate the nature and the concentration of paramagnetic species.

## GLASS PREPARATION AND CHARACTERIZATION METHODS

Two series of glasses were prepared for this study all derived from the following composition (mol.%): $58.2SiO_2$ - $13.77Na_2O$ - $9.81CaO$ - $18.08B_2O_3$ either by increasing $MoO_3$ concentration from 0 to 5.0 ($M_x$ series with x = 0, 0.87, 1.54, 2.50, 3.62 and 5 mol.% $MoO_3$) or by changing $B_2O_3$ concentration from 0 to 24 mol.% ($B_y$ series with y = 0, 6, 12, 18 and 24 mol.% $B_2O_3$) keeping constant $MoO_3$ concentration (2.50 mol.%). For all samples, 0.15 mol.% $Nd_2O_3$ was introduced in composition both to facilitate $^{29}Si$ nuclei relaxation during MAS NMR experiments and to perform optical studies not presented in this paper [4]. Glasses were all prepared at 1300°C under air in Pt crucibles using reagent grade $SiO_2$, $CaCO_3$, $Na_2CO_3$, $H_3BO_3$, $MoO_3$ and $Nd_2O_3$ powders. Depending on glass composition, samples were quenched either as cylinders or disks [4]. Several reference glass samples (borate and silicate glasses) were also prepared for comparison with $M_x$ and $B_y$ glasses (NMR and Raman spectra). The amorphous character of samples was checked using both X-ray diffraction

(XRD) and Raman spectroscopy. Unpolarized Raman spectra of monolithic samples were collected with T64000 Jobin-Yvon confocal Raman spectrometer operating at approximately 1.5 W at room temperature with the 488 nm line of an argon ion laser for excitation. $^{29}$Si MAS NMR spectra were recorded on a Bruker Avance 300 spectrometer operating at 59.63 MHz. $^{11}$B MAS NMR spectra were recorded on a Bruker Avance 400 operating at 128.28 MHz. $^{23}$Na MAS NMR spectra were recorded on a Bruker Avance II 500WB spectrometer operating at 132.03 MHz. Chemical shifts were determined relative to tetramethylsilane for $^{29}$Si, liquid $BF_3OEt_2$ for $^{11}$B and 1.0M aqueous NaCl solution for $^{23}$Na. ESR spectra were recorded on a Bruker ELEXYS E500 spectrometer operating at X band (9.5 GHz) in the range of temperature 20-300 K. For all glasses of $M_x$ and $B_y$ series, ESR showed the existence of a signal due to Mo near g~1.91 and that can be detected at least from 20K to room temperature. These ESR characteristics indicated that this signal is due to paramagnetic $Mo^{5+}$ ($4d^1$) ions located in low symmetry sites. Indeed, the spin-lattice relaxation time of $d^1$ ions is known to increase (and thus the possibility to detect the ESR signal at high temperature also) with the distortion of the sites. This result is in agreement with the paper of Farges et al. [5] which proposed that the ESR signal of Mo in glasses was associated with low symmetry molyddenyl entities. No signal associated with $Mo^{3+}$ ($4d^3$) ions near g~5.19 was detected on ESR spectra [5]. For instance, at 20K only a low intensity contribution due to $Nd^{3+}$ and $Fe^{3+}$ (impurity) ions was detected in the low field region of the spectra. The proportion of $Mo^{5+}$ ions (over all molybdenum) ranges between 0.4 and 0.8 % for all the glasses studied in this work as estimated using a DPPH sample as concentration standard. Consequently, the majority of molybdenum (> 99%) occurs as $Mo^{6+}$ ions in glasses of $M_x$ and $B_y$ series prepared under air (oxidizing conditions). According to Mo EXAFS and XANES results in silicate glasses and to bond valence-bond length considerations published in literature, $Mo^{6+}$ ions are present as tetrahedral molybdate entities $MoO_4^{2-}$ in modifiers rich regions of the glass structure (depolymerized regions) and are not linked directly to the silicate network [1,5,6].

## STRUCTURAL EVOLUTION OF GLASSES WITH INCREASING MoO$_3$ CONCENTRATION

Raman spectra confirm the XRD results presented in [4] showing that the solubilty limit of molybdenum in $M_x$ glasses was reached between 1.54 and 2.5 $MoO_3$ mol.%. Indeed, Fig. 1 clearly reveals the occurrence of the contribution of $CaMoO_4$ (powellite) Raman vibration modes for x > 1.54 mol.%. For comparison, the Raman spectrum of a powellite ceramic sample is given with the attribution of the bands according to [7]. All the $CaMoO_4$ bands with frequency ≥ 321 cm$^{-1}$ correspond to internal vibrational modes of $MoO_4^{2-}$ tetrahedra and the strongest band at 879 cm$^{-1}$ can be associated with the symmetric streching vibration of Mo-O bonds. By analogy, we propose that the wide and intense band observed in the 898-913 cm$^{-1}$ range on the Raman spectra of all glasses of $M_x$ series (and also of the $B_y$ series) is also due to the symmetric streching vibration of Mo-O bonds of molybdate tetraedra within the glass structure. Fig. 2 indicates that this band moves towards lower frequencies when x increases (x ≥ 2.5) which shows that the environnment and/or the symmetry of $MoO_4^{2-}$ tetrahedra in the glass is modified at least when the crystallization of powellite is detected. Comparison of $M_x$ spectra with the spectrum of a glass without $Ca^{2+}$ ions and belonging to the $SiO_2$-$Na_2O$-$MoO_3$ system ($B_0$(Na) glass in Fig. 2) seems to indicate that the amount of $Na^+$ ions acting as charge compensators near $MoO_4^{2-}$ tetrahedra increases with x at the expenses of $Ca^{2+}$ ions. This evolution can be explained by the increase of the Na/Ca ratio in the modifiers-rich regions of the glass structure when powellite is formed. Thus, Raman spectroscopy of glasses containing Mo seems to be more sensisitive than EXAFS to detect local composition variations around $MoO_4^{2-}$ tetrahedra (and thus symmetry modifications) in the glass structure. Indeed, the Mo

EXAFS results published in literature gave very similar Mo-O distance for different silicate glass compositions (1.76-1.78 Å) [1,5,6].

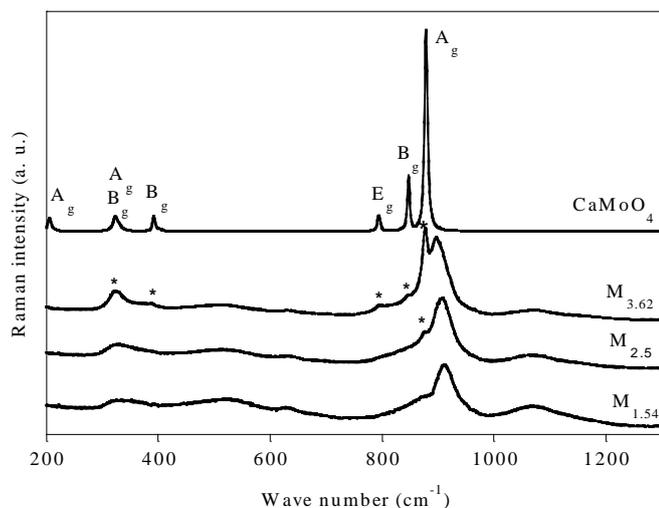

**Fig. 1**. Normalized Raman spectra of $M_{1.54}$, $M_{2.5}$ and $M_{3.62}$ glasses. The Raman spectrum of a $CaMoO_4$ (powellite) ceramic is given for comparison. Spectra were not corrected with the Long formula. *: vibration bands due to $CaMoO_4$ crystals in $M_x$ samples.

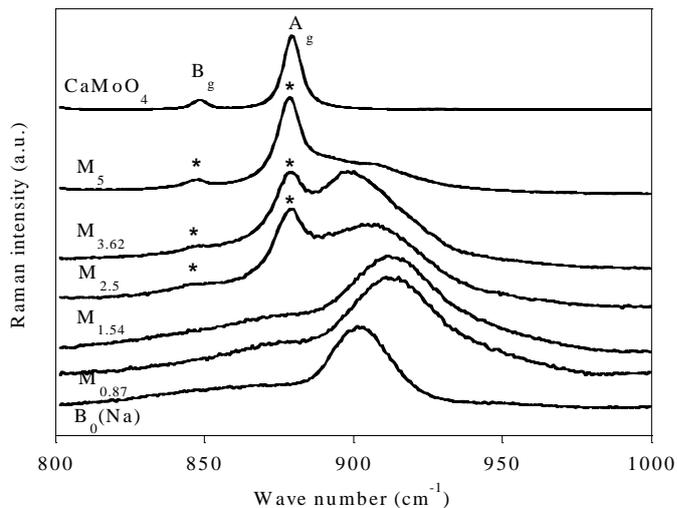

**Fig. 2**. Normalized Raman spectra of $M_{0.87}$, $M_{1.54}$, $M_{2.5}$, $M_{3.62}$ and $M_5$ glasses. The spectra of a $CaMoO_4$ (powellite) ceramic and of sodium silicate glass with Mo ($69.34SiO_2 - 28.09Na_2O - 2.43MoO_3 - 0.15Nd_2O_3$ in mol.%) are given for comparison. *: vibration bands due to $CaMoO_4$ crystals in $M_x$ samples.

$^{29}Si$ MAS NMR spectra were simulated with three bands centered at -80.0, -92.2 and -103.6 ppm respectively associated with $Q_2$, $Q_3$ and $Q_4$ units ($Q_n$ units with n = 0 to 4 correspond to $SiO_4$ tetrahedra with n bridging oxygen atoms). These chemical shift values were kept constant for the simulation of the spectra of all samples of $M_x$ and $B_y$ series. An example of curve-fitting is shown in Fig. 3a and the evolution of the relative proportions [$Q_n$] of $Q_n$ units is shown in Fig. 3b. This evolution reveals that [$Q_2$] and [$Q_3$] decrease whereas [$Q_4$] increases when molybdenum concentration increases in samples of the $M_x$ series: when $MoO_3$ increases from 0 to 5 mol.%, the proportion of $Q_4$ units increases of more than 20 % (Table 1).

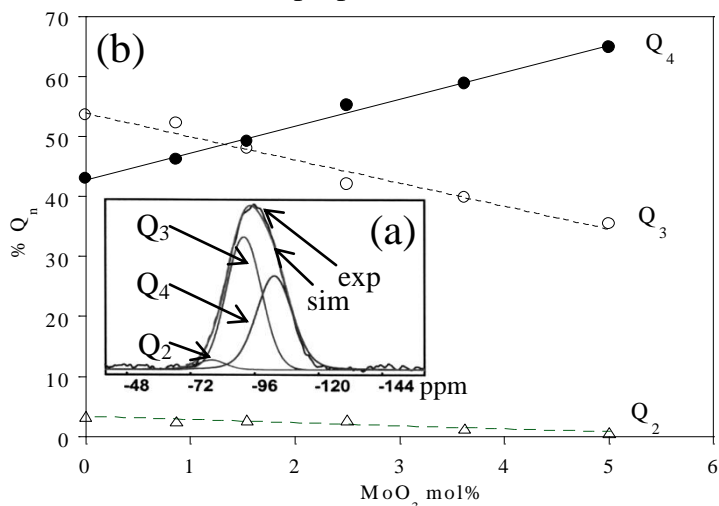

**Fig. 3**. (a) Example of $^{29}Si$ MAS NMR spectra recorded for the $M_0$ sample. The corresponding simulation using three Gaussian line shape contributions associated with $Q_2$, $Q_3$ and $Q_4$ units is shown (exp: experimental spectrum, sim: simulated spectrum). The same chemical shift values were used for the spectra simulation of all the samples of $M_x$ and $B_y$ series. (b) Evolution of the relative proportions of $Q_4$, $Q_3$ and $Q_2$ units in $M_x$ samples with the increase of $MoO_3$ concentration. Linear fits of $Q_n$ evolution are shown.

For the $M_x$ series, $^{11}B$ MAS NMR spectra simulation only shows a slight and non-monotonous decrease of the relative proportion of $BO_4^-$ units when molybdenum concentration increases: the variation of the proportion of $BO_4^-$ units was only about 2-4 % (Table 1). Consequently, $MoO_3$ acts as a reticulating agent for the silicate network in $M_x$ glasses and $MoO_3$ mainly acts on the amount of $Q_3$ units (Table 1). This result can be explained as follows. As molybdenum is introduced as $MoO_3$ (corresponding to one $Mo^{6+}$ ion

|  | $M_0$ | $M_{0.87}$ | $M_{1.54}$ | $M_{2.50}$ | $M_{3.62}$ | $M_5$ |
|---|---|---|---|---|---|---|
| % $Q_4$ | 43 | 46.2 | 49.2 | 55.2 | 58.8 | 64.8 |
| % $Q_3$ | 53.6 | 52.2 | 48.0 | 42.0 | 39.8 | 34.5 |
| % $Q_2$ | 3.4 | 2.6 | 2.8 | 2.8 | 1.4 | 0.7 |
| $n_{Q3}$ | 31.19 | 30.38 | 27.93 | 24.44 | 23.16 | 20.08 |
| $n_{Mo}$ | 0 | 0.87 | 1.56 | 2.56 | 3.75 | 5.26 |
| $\Delta n_{Q3}$ | - | 0.81 | 3.26 | 6.75 | 8.03 | 11.11 |
| $2n_{Mo}$ | 0 | 1.74 | 3.12 | 5.12 | 7.5 | 10.52 |
| % $BO_3$ | 46.0 | 43.8 | 46.4 | 47.8 | 49.7 | 47.8 |
| % $BO_4$ | 54.0 | 56.2 | 53.6 | 52.3 | 50.3 | 52.3 |
| $[BO_4]/[BO_3]$ | 1.17 | 1.28 | 1.15 | 1.09 | 1.01 | 1.09 |

**Table 1**. Relative proportions of $Q_n$ units (n = 2, 3, 4) and ($BO_3$, $BO_4^-$) units in $M_x$ samples determined after simulation and integration of $^{29}Si$ and $^{11}B$ MAS-MNR spectra respectively. For a constant number of moles of $SiO_2$ (58.2 in $M_0$ composition), the number of moles of $Mo^{6+}$ ions ($n_{Mo}$) and $Q_3$ units ($n_{Q3}$) is reported for $M_x$ samples. The number of moles of $Q_3$ units that disappeared ($\Delta n_{Q3}$) when x increased (in comparison with $M_0$ glass) is also reported.

and 3 non-bridging atoms of oxygen (NBO)) in glass batch whereas $Mo^{6+}$ ions are known to occur as $MoO_4^{2-}$ units (corresponding to one $Mo^{6+}$ ion and 4 NBO) both in glass structure and powellite crystals, each $Mo^{6+}$ ion needs to catch one NBO more from the borosilicate network. We thus propose the following reaction scheme between $MoO_3$ and $Q_3$ units (initially charge compensated by $Na^+$ or $Ca^{2+}$ ions) in the melt:

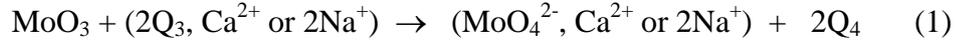
$$MoO_3 + (2Q_3, Ca^{2+} \text{ or } 2Na^+) \rightarrow (MoO_4^{2-}, Ca^{2+} \text{ or } 2Na^+) + 2Q_4 \quad (1)$$

For a constant number of moles of $SiO_2$ (58.2 in $M_0$ composition), the number of moles of $Mo^{6+}$ ions ($n_{Mo}$) and $Q_3$ units ($n_{Q3}$) was calculated for all $M_x$ samples and is reported in Table 1. The comparison of $\Delta n_{Q3}$ (the number of moles of $Q_3$ units that have disappeared in $M_x$ sample in comparison with $M_0$ sample) with $2n_{Mo}$ (see equation (1)) shows that the values of $\Delta n_{Q3}$ and $2n_{Mo}$ remain close to each other when the amount of $MoO_3$ increases in glass composition which seems to confirm the reaction scheme (1) proposed above.

## STRUCTURAL EVOLUTION OF GLASSES WITH INCREASING $B_2O_3$ CONCENTRATION

In [4] we showed that $Na_2MoO_4$ crystallization tendency during slow cooling of the melt (1°C/min) decreased with the increase of $B_2O_3$ concentration whereas the tendency of $CaMoO_4$ to crystallize increased. Such as evolution can be explain by the preferential charge compensation of $BO_4^-$ units by $Na^+$ rather than by $Ca^{2+}$ ions in borosilicate glasses [8]. For the $B_y$ series, Fig. 4 shows that the $[BO_4^-]/[SiO_2]$ ratio increases whereas $[Na^+]/[BO_4^-]$ decreases with $B_2O_3$ concentration. It is interesting to notice that for the $B_{24}$ sample almost all $Na^+$ ions can act as $BO_4^-$ charge compensator ($[Na^+]/[BO_4^-] \sim 1$). In these conditions, the amount of $Na^+$ ions able to compensate the $MoO_4^{2-}$ entities strongly decreases when $B_2O_3$ concentration increases and the $[Ca^{2+}]/[Na^+]$ ratio in the depolymerized regions of glass structure increases which can explained the evolution of the crystallization tendency. Fig. 5 shows that the isotropic $^{23}Na$ chemical shift ($\delta_{iso}(^{23}Na)$) decreases when $B_2O_3$ concentration increases. Thus, the distribution of $Na^+$ ions through the glassy network significantly changes when increasing amounts of boron are introduced in $B_y$ glasses. Comparison of $\delta_{iso}(^{23}Na)$ of $B_y$ glasses with that of sodium silicate (SiNa), sodium calcium silicate (SiNaCa) and borate (B0.2Na, B0.7Na) reference glasses clearly reveals that when $B_2O_3$ concentration increases, $Na^+$ ions moves from a charge compensator position near NBO to a charge compensator position near $BO_4^-$ units.

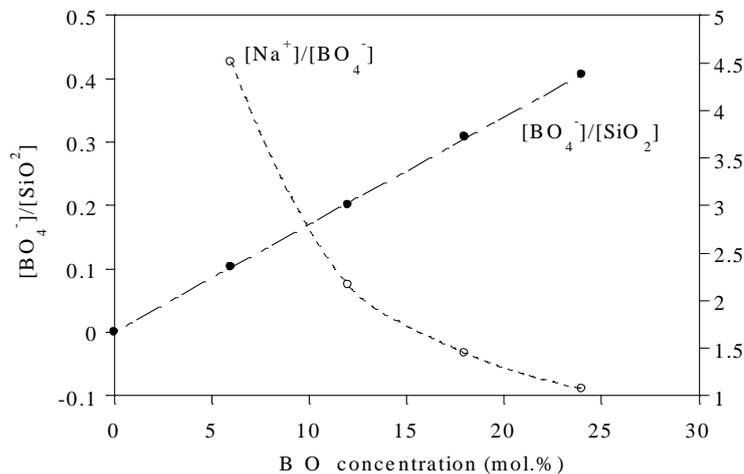

**Fig. 4.** Evolution of the $[BO_4^-]/[SiO_2]$ and $[Na^+]/[BO_4^-]$ ratios versus $B_2O_3$ concentration in $B_y$ samples (mol.%). The $Na^+$ and $SiO_2$ concentrations were determined by chemical analysis whereas the $BO_4^-$ concentration was determined by chemical analysis and $^{11}B$ MAS NMR.

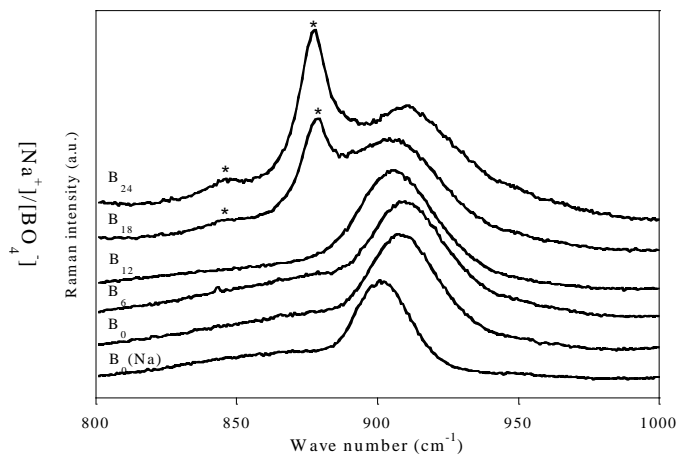

**Fig. 6.** Evolution of Raman spectra of $B_y$ samples. For comparison the spectrum of the $B_0(Na)$ reference glass without calcium is also shown. *: vibration bands due to $CaMoO_4$ crystals in $M_x$ samples.

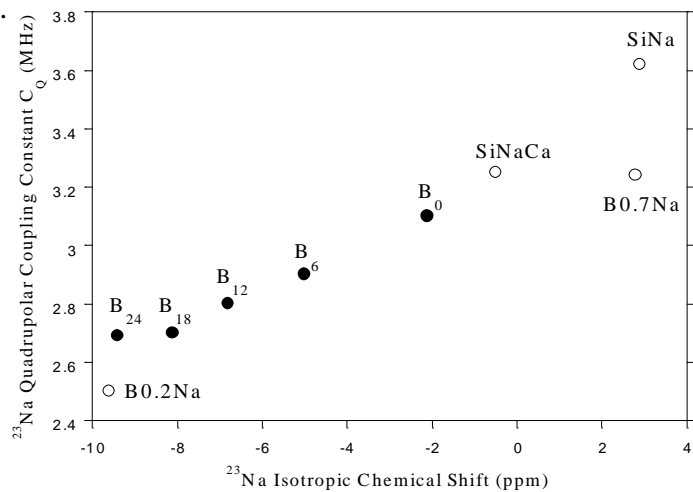

**Fig. 5.** Evolution of the $^{23}Na$ isotropic chemical shift ($\delta_{iso}$) and quadrupolar coupling constant ($C_Q$) in the samples of $B_y$ series. For comparison the values of $\delta_{iso}$ and $C_Q$ of reference glasses are also shown: SiNa ($80.93SiO_2 - 19.07Na_2O$), SiNaCa ($71.21SiO_2-16.78Na_2O-12CaO$), B0.7Na ($58.8B_2O_3-41.2Na_2O$), B0.2Na ($83.3B_2O_3-16.7Na_2O$). For the three former reference glasses $Na^+$ ions can compensate NBO whereas in the later one $Na^+$ ions only compensate bridging oxygen atoms near $BO_4^-$ units.

In accordance with the XRD results on the $B_y$ quenched disk samples, Raman spectra show that the crystallization of $CaMoO_4$ is detected when $B_2O_3$ concentration is higher than 12 mol.% (Fig. 6). Contrary to the Raman spectra of the samples of the $M_x$ series, the position of the band associated with Mo-O streching vibration near 905 cm$^{-1}$ only slightly evolutes when $B_2O_3$ concentration increases which indicates that the environment of $MoO_4^{2-}$ entities is only slightly modified. As the depolymerized regions in which are located $MoO_4^{2-}$ entities become progressively depleted in sodium when $B_2O_3$ concentration increases, the lack of strong evolution of the M-O vibrationnal frequency could indicate that $MoO_4^{2-}$ entities are preferentially charge compensated by $Ca^{2+}$ ions.